\documentclass[10pt]{IEEEtran}
\usepackage[dvipdfm]{color}
\usepackage[dvipdfm]{graphicx}
\usepackage{amsmath}
\usepackage{amsfonts}
\usepackage{amssymb}

\AtBeginDvi{}

\title{\LARGE \bf Scaling properties of complex networks: \\
Toward Wilsonian renormalization for complex networks}

\date{\today}
\author{
 \IEEEauthorblockN{
 Kento Ichikawa\IEEEauthorrefmark{1},
 Masato Uchida\IEEEauthorrefmark{2},
 Masato Tsuru\IEEEauthorrefmark{3},
 Yuji Oie\IEEEauthorrefmark{4},
 }\\
 \IEEEauthorblockA{
 \IEEEauthorrefmark{1}
 Department of Computer Science and Systems Engineering,\\
 Kyushu Institute of Technology \\
 Email: ichikawa@ndrc.kyutech.ac.jp \\
 }
 \IEEEauthorblockA{
 \IEEEauthorrefmark{2}
 Network Design Research Center, Kyushu Institute of Technology \\
 Email: m.uchida@ndrc.kyutech.ac.jp \\
 }
 \IEEEauthorblockA{
 \IEEEauthorrefmark{3}
 Department of Computer Science and Systems Engineering,\\
 Kyushu Institute of Technology \\
 Email: tsuru@cse.kyutech.ac.jp \\
 }
 \IEEEauthorblockA{
 \IEEEauthorrefmark{1}
 Department of Computer Science and Systems Engineering,\\
 Kyushu Institute of Technology \\
 Email: oie@cse.kyutech.ac.jp \\
 }
}

\begin{document}

\maketitle

\begin{abstract}
 

Nowadays, scaling methods for general large-scale complex networks have
been developed.  We proposed a new scaling scheme called ``two-site
scaling''. This scheme was applied iteratively to various networks, and
we observed how the degree distribution of the network changes by two-site
scaling. In particular, networks constructed by the BA algorithm behave
differently from the networks observed in the nature. In addition,
an iterative scaling scheme can define a new renormalizing method. We
investigated the possibility of defining the Wilsonian renormalization group
method on general complex networks and its application to the analysis 
of the dynamics of complex networks.
\end{abstract}

\section{Introduction}


In recent years, complex networks have been actively
investigated. Thanks to recent increases in computational power, we can
deal with huge and complex networks that exist in the real world. One
subject of particular interest is the cross-disciplinary nature of
complex networks. For example, as pointed out by Barabasi et al., the
cinema actors' costarring network and the Internet to the metabolic
network of cells have power-law or scale-free properties
\cite{Barabasi1999,Albert2002}. To explain the scale-free property, the
BA algorithm\cite{Barabasi1999a} has been proposed. The BA algorithm is
a simple growing network model based on the notion of preferential
attachment.


Other than frequently discussed quantities such as the powers of the
power-law distributions and the cluster coefficient, we can define
numerous quantities characterizing a network, but some of these
quantities are difficult to calculate because of computational
complexity, and we scarcely know which quantities should be used in the
study of complex networks for the realistic applications.


On the other hand, recent studies on complex networks have focused on
the dynamics of the network\cite{Boccaletti2006}.  It is desirable to
understand the dynamics of the network from an application point of
view, because there are many demands related to the control of the
dynamics of various networks. However, these studies depend on the
individual dynamic system and the algorithms used to generate complex
networks. It is necessary to take a systematic approach to the dynamics
of complex networks.

In the present paper, we propose a scaling method called two-site
scaling. As mentioned in the following section, scaling is essential for
the Wilsonian renormalization method. Using the Wilsonian
renormalization method, we can extract information about dynamics, such
as critical exponents, systematically, which is expected to by universal
in nature and would enables us to classify the dynamics on networks and
underlying networks from a dynamics point of view.

\section{Wilsonian renormalization group method}


Wilsonian renormalization methods\cite{Cardy1996,Benny1992} are powerful
and systematic methods in theoretical physics.  The Wilsonian
renormalization theory is the theory of the flow of the parameters in
the parameter space of dynamic systems. Therefore, we can expect the
renormalization method to be a systematic approach to the dynamics of
the complex networks.  In particular, in condensed matter physics,
critical exponents derived by the renormalization method are known
empirically to have strong ``universality''\cite{Stanley1999,Benny1992},
which requires that the systems have same dimension and that the numbers
of states have the same critical exponent.  The renormalization method
usually deploys scaling of lattices.


Here, there is another important procedure, referred to as 
rescaling, in the Wilsonian renormalization theory. This is a 
procedure by which to integrate the dynamics in the subgraphs, which is contracted
by scaling, and represent their contributions by renormalizing the
parameters of the dynamics, for example, parameters of the
Hamiltonian. However, in the present paper, we do not discuss rescaling, but
rather focus only on how to define the scaling in complex networks.


Thanks to the universality, the critical exponents of the systems are 
good measures for classifying dynamic systems on the
networks. These classes are called universality classes. For example,
critical exponents for Ising models can be derived numerically by using,
for example, the Metropolis method. 

Renormalization of some network structural quantities has already been
applied to complex networks such as Watts-Strogatz's small world
network\cite{Newman1999} and the geographical embedded
network\cite{BJKim2004}. These scalings can be defined naturally because
these networks are embedded in Euclidean space, and scaling can be
defined naturally by the scaling of the Euclidean space. However, in
these papers only structural quantities are renormalized. Thus, they are
not Wilsonian renormalization procedures.


In the general complex network, there is no such naturally obtained
scaling method. Therefore, we should define how to scale these
networks. In recent years, scaling methods have been proposed for
general complex networks\cite{Song2005,JSKim2007}. We are prepared to
define renormalization group method for general complex networks.


We define the action of the renormalization group by the two-site
scaling method defined in Section~\ref{sec:twosite}. Note that, as
mentioned in Section~\ref{sec:renorm}, further study is needed in order
to verify that the proposed two-site scaling method will yield an
appropriate renormalization scheme. In Section~\ref{sec:scaling}, we
apply the proposed two-site scaling method iteratively to various
networks, including the BA network, the Internet router network, the AS
network, the network of actors' costars, and the protein-protein
interaction network.

\section{Two-site scaling}
\label{sec:twosite}


The scaling method for general complex networks used herein is the box
covering method, which was proposed by Song-Havlin-Makse\cite{Song2005}
and investigated in successive studies \cite{JSKim2007}. The box
covering method divides a network to subgraphs of diameter smaller than
the given length. Thus, we can regard these methods as coarse
grained. In the terminology of graph theory, this procedure corresponds
to contraction\footnote{In previous papers, only the number of subgraphs
was counted, so that this graph theoretical interpretation was not
necessary}. Contraction means that the subgraphs are considered to be
one node $v$ and the links between the inside of the subgraph and the
outside of the subgraph are considered to be a link to the node $v$. In
the scaling, we divide the network into subgraphs called boxes and
contract these boxes.



Two-site scaling is a method of dividing a network into randomly
selected pairs of adjacent nodes and contracting these pairs. However,
if networks have many degree-1 nodes (referred to herein as leaves),
randomly selected pairs tend to be pairs that consist of a leaf and the
adjacent node, and contraction is reduced to simply removing one leaf.
In the Wilsonian renormalization theory, the information to be extracted
from the object systems is important. Since we intend to perform
two-site scaling homogeneously over the network and to extract
information about the dynamics and the network homogeneously over the
entire network, this is inappropriate because it makes the two-site
scaling method inhomogeneous in the networks. To avoid this, we propose
ignoring leaves and apply this procedure to nodes of degree greater than
or equal to 2. For degree-1 nodes (leaves), pairs of leaves having
adjacent nodes that are identical are contracted.


\begin{figure}[t]
 \begin{center}
  \includegraphics[scale=0.5]{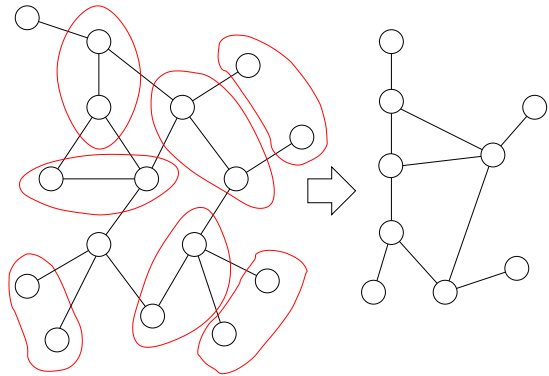}  
 \end{center}
 \caption{Example of two-site scaling}
 \label{fig:two-site}
\end{figure}

Summarizing the above, we have following algorithm.
\begin{enumerate}
 \item Count adjacent degree-1 nodes for nodes of degree greater 
 than 1
 \item Remove degree-1 nodes
 \item Fill the network with randomly selected pairs of adjacent nodes
 \item Contract the pairs constructed in Step 2.
 \item Add half of the number of degree-1 nodes
\end{enumerate}


By this algorithm, the sizes (number of nodes) of the networks without
leaves are approximately halved. We can iterate this algorithm and scale
networks in a systematic manner. Note that previous scaling methods do
not consider this iteration.

\subsection{Comparison with other scaling methods}


The method defined above resembles renormalization by block spin
transformation (BST) in condensed matter physics. For a $d$ dimensional
lattice, the usual method is to divide the lattice into $d$-dimensional
cubes and contract the $d$-dimensional cube as one site.


However, since in general complex networks, we cannot expect
$d$-dimensional cubes to appear homogeneously, we need another method.
The box covering method by \cite{Song2005} divides a network into boxes,
where any distance between any pair of nodes in a box is less then
$\ell_S$. In a simplified version of this method by \cite{Song2005,JSKim2007}, a
node $v$ is selected randomly, and nodes located a distance from $v$ of less
than $\ell_B$ constitute a box.



The box covering method resembles to the scaling method used to
calculate the fractal dimension. The intent is not to define the
renormalization scheme, but rather to calculate the fractal dimension of
complex networks. As a result, there are some problems, such as
dispersion of the box sizes, if we employ the box covering method as a
scaling method for the renormalization.

The other methods by which to define the renormalization scheme are to
embed complex networks into a 1-dimensional lattice or a $d$-dimensional
lattice or to contract a $d$-dimensional
cube\cite{Newman1999,Rozenfeld2002,benAvraham2003,BJKim2004}. These
methods can be extended to the natural extension of renormalization but
cannot adapt to general complex networks.


\subsection{Features of two-site scaling}


One characteristic of two-site scaling is that one iteration of two-site
scaling is homogeneous contraction, i.e., almost all nodes are included
in boxes of the same size and the size of each box is two, which is the
minimum number for contraction.  This characteristic also minimizes the
loss of the fine link structure. Therefore, we can have a coarse-grained
network of the required size with less loss of the fine structure of the
network.
 

On the other hand, this may become problematic, as explained above,
because special treatment is required for nodes of degree 1. Without
this special treatment, we cannot scale the network homogeneously.

\section{Scaling of degree distributions of various networks}
\label{sec:scaling}

\begin{figure}[t]
 \begin{center}
  \includegraphics[scale=0.5]{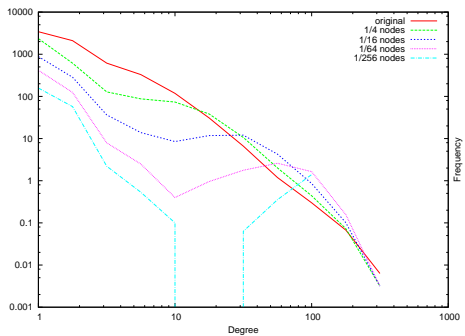}
 \end{center}
 \caption{Scaling of the network by the BA algorithm: scaling property of the degree
 distribution by every two iterations of two-site scaling}
 \label{fig:scale-ba}
\end{figure}
\begin{figure}
 \begin{center}
  \includegraphics[scale=0.5]{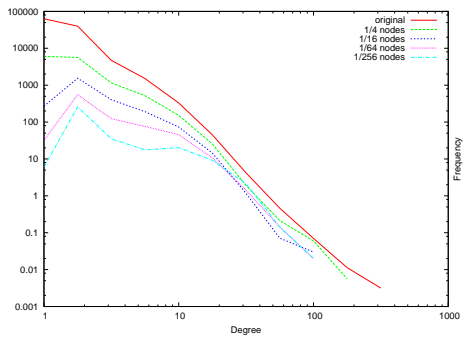}
 \end{center}
 \caption{Scaling of the router network of the Internet}
 \label{fig:scale-rt}
\end{figure}
\begin{figure}[t]
 \begin{center}
  \includegraphics[scale=.5]{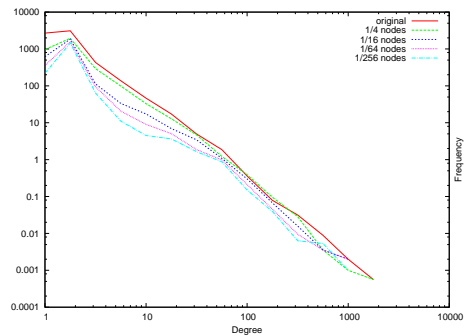}
 \end{center}
 \caption{Scaling of the AS network of the Internet}
 \label{fig:scale-as}
\end{figure}
\begin{figure}[t]
 \begin{center}
  \includegraphics[scale=.5]{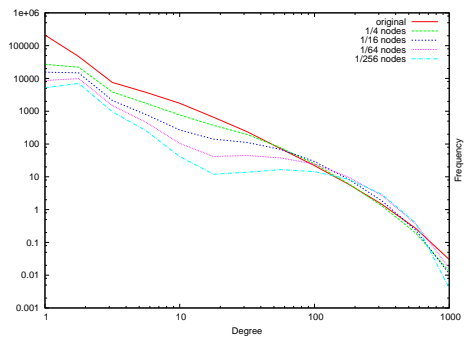}
 \end{center}
 \caption{Scaling of the actors' costarring network}
 \label{fig:scale-actor}
\end{figure}
\begin{figure}
 \begin{center}
  \includegraphics[scale=.5]{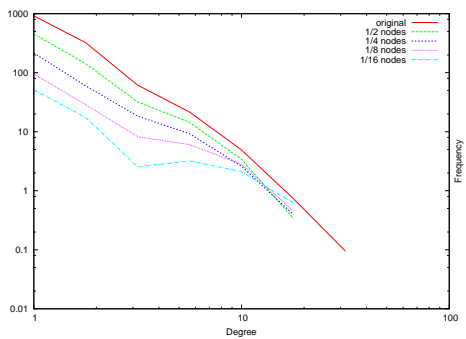}
 \end{center}
 \caption{Scaling of the protein-protein interaction network}
 \label{fig:scale-protain}
\end{figure}


Figure~\ref{fig:scale-ba}, Figure~\ref{fig:scale-rt},
Figure~\ref{fig:scale-as}, Figure~\ref{fig:scale-actor},
and Figure~\ref{fig:scale-protain} show the scaling property of the degree
distribution of the network constructed by the BA algorithm, the Internet
topologies of the router level and the AS level, the actor's costarring network, and
protein-protein interaction network, respectively. The two-site scaling
method acts iteratively on these networks and the degree distributions of
these networks are plotted for every two (one for
protein-protein networks) iterations.

As shown in Figure~\ref{fig:scale-ba}, the degree distribution of the network
constructed by the BA algorithm shows rapid transition from a power-law
distribution to a bimodal distribution. On the other hand, other networks
observed in nature maintain a power law distribution.




\section{Discussion}


The two-site scaling method, which is inspired from block spin
transformation in renormalization theory, assigns new quantities to
complex networks. We would like to emphasize that the quantities given
by scaling methods would influence the dynamics of networks, especially
with respect to critical phenomena. Although we could not calculate the
critical exponents in the present study, these critical exponents are
expected to have a universal property, and would be used to classify the
universal classes of dynamic systems on complex networks.

\subsection{Peculiarity of BA networks}

We found that the degree distribution of networks constructed by the BA
algorithm is transformed from a power-law distribution into a bimodal
distribution. This is unusual compared to networks observed in nature,
such as the Internet, the actor network, and the protein-protein
interaction network, which maintain a power law distribution by the
iteration of two-site scaling. In other words, networks observed in
nature would have more strong scale invariance in terms of not only the
degree distribution, but also the proposed scaling method.

The reason for this peculiarity of BA
networks remains unknown. This phenomenon should be analyzed in a future study.

\subsection{Toward the Wilsonian renormalization group method}
\label{sec:renorm}


Studies of complex networks are often accompanied by the problem of
computational complexity because many elemental problems, such as
the isomorphism problem, are known to be NP\cite{Garey1979}. Argument by
the renormalization method with scaling of the complex network would provide
a scalable theory of the dynamics of complex networks.


The two-site scaling method resembles block spin transformation in
condensed matter physics. There have been attempts to define the fractal
dimension of complex networks\cite{Song2007}. However, the dimension of
complex networks cannot be defined in a straightforward
manner. Therefore, further investigation is needed in order to verify
that the two-site scaling method can define the renormalization group
action.


The renormalization scheme defined from the box covering method, which
was originally proposed to define the fractal dimension, would be robust
with respect to differences in fractal dimension. However, since the
iteration of the box covering rapidly contracts a large network to one
node, it is difficult to obtain information such as critical exponents
from the few iterations of the box covering method.  This is because the
computation of renormalization should be executed numerically, and the
numerical error of critical exponents becomes large.

In future studies, we intend to verify that the renormalization scheme
defined by the two-site scaling function is similar to other block spin
transformations. This is a difficult problem in that the proposed
scaling scheme maps a regular lattice not to a regular lattice, but to
complex networks. For this and other reasons, we should employ numerical
methods such as Monte Carlo Renormalization. It may be useful to verify
that this problem will be overcome by applying this scheme to regular
lattices, the universality classes of which are determined. If we have
the same critical exponent from two-site scaling method as other
renormalization schemes, we can expect these exponents to provide the
classification of dynamics and underlying networks.

Moreover, since networks scaled by the BA algorithm have peculiar
scaling properties, as compared with natural networks, BA networks have
a different structure than natural networks, and this difference may
affect the dynamics on BA networks, especially for critical phenomena.

\section{Conclusion}

We have proposed a scaling method called two-site scaling. This scaling
scheme maintains the shape of the degree distribution of various natural
networks. However, the degree distribution of BA networks is changed
drastically by this scaling method. Networks observed in nature maintain
the shapes of their degree distributions by the proposed scaling
methods, which means that these networks have stronger scale invariance
than scale-free degree distributions. Finally, we discussed the
application of the proposed scaling method to the Wilsonian
renormalization method. A great deal of research remains to be conducted
in order to complete the Wilsonian renormalization theory for complex
networks.

Note added: In this paper we proposed a roadmap to the Wilsonian
renormalization group method, but we informed that a preprint by
F. Radicchi et. el.~\cite{Radicchi2008a} has been submitted to preprint
server just before the presentation on the PHSYCOMNET workshop, but
after the acceptance ( Dec.1st 2007 ) of this paper.  Although the paper
by Radicchi et. el. employs greedy coloring algorithm(GCA) and random
burning(RB) for box covering method, insists a similar roadmap to the
Wilsonian renormalization group method.

\section*{Acknowledgments}
The authors would like to thank Naoki Masuda for his helpful comments
and advice.  This work was supported in part by the Ministry of Internal
Affairs and Communications, Japan and by the Japan Society for the
Promotion of Science through a Grant-in-Aid for Scientific Research
(S)(18100001).

\bibliographystyle{IEEEtran}
\bibliography{renorm}
\end{document}